\documentclass[%
 reprint, amsmath,amssymb,aps,prl,longbibliography,lengthcheck,%
]{revtex4-1}

\usepackage{graphicx}
\usepackage{dcolumn}
\usepackage{bm}
\usepackage{hyperref}
\usepackage{float}
\usepackage{pstricks}
\usepackage{axodraw}

\newcommand{\en}{\enspace}
\newcommand{\mfzsixh}{ 0.478 }
\newcommand{\Gfzsixh}{ 0.324 }
\newcommand{\fz}[1]{f_0( #1 )}
\newcommand{\Ff}[1]{F_{f_0( #1 )}}
\newcommand{\wQ}{\cal Q}

\begin{document}

\preprint{}

\title{Tetraquark interpretation of the Belle data on the anomalous
 $\Upsilon(1S) \pi^+\pi^-$ and $\Upsilon(2S) \pi^+\pi^-$ production near the
$\Upsilon(5S)$ resonance} 

\begin{figure}[H]
\fcolorbox{white}{white}{
  \begin{picture}(500,16) (0,0)
  \Text(476,9)[b]{DESY 09-222}
\Text(473,0)[b]{December 2009}
  \end{picture}
  }
\end{figure}

\author{Ahmed~Ali}
\email{ahmed.ali@desy.de}
\author{Christian Hambrock}
\email{christian.hambrock@desy.de}
\affiliation{Deutsches Elektronen-Synchrotron DESY, D-22607 Hamburg, Germany}

\author{M. Jamil Aslam}
\email{muhammadjamil.aslam@gmail.com}
\affiliation{Physics Department, Quaid-i-Azam, University, Islamabad,
 Pakistan}

\date{\today}

\begin{abstract}
We analyze the Belle data [K.~F.~Chen {\it et al.} (Belle Collaboration),
Phys.\ Rev.\ Lett.\  {\bf 100}, 112001 (2008);
I.~Adachi {\it et al.}  (Belle Collaboration), arXiv:0808.2445] on the
 processes 
$e^+ e^- \to \Upsilon(1S)\; \pi^+\pi^-, \Upsilon(2S)\; \pi^+\pi^-$ near the
peak of the $\Upsilon(5S)$ resonance, which are found to be anomalously
large in rates compared to similar dipion
transitions between the lower $\Upsilon$ resonances.
Assuming these final states arise from the production and decays of
the $J^{PC}=1^{--}$ state $Y_b(10890)$, which we interpret as a
bound (diquark-antidiquark) tetraquark state
 $[bq][\bar{b}\bar{q}]$, a dynamical model for the  decays
$Y_b \to  \Upsilon(1S)\; \pi^+\pi^-,  \Upsilon(2S)\; \pi^+\pi^-$
is presented. Depending on the phase space, these decays 
 receive significant contributions
from the scalar $0^{++}$ states, $f_0(600)$ and $f_0(980)$, and from the
$2^{++}$ $q\bar{q}$-meson $f_2(1270)$.
Our model provides excellent fits for the decay distributions,
 supporting $Y_b$ as a tetraquark state. 
\end{abstract}


\maketitle
The observation of the $\Upsilon(1S)\;\pi^+\pi^-$ and $\Upsilon(2S)\;\pi^+\pi^-$ states near the 
$\Upsilon(5S)$ resonance peak at $\sqrt{s}=10.87$ GeV at the KEKB $e^+e^-$ collider
by the Belle collaboration~\cite{Abe:2007tk} has received a lot of theoretical 
attention~\cite{Hou:2006it}.
The two puzzling features of these data are that, if interpreted in terms of the
processes
$e^+ e^- \to \Upsilon(5S) \to \Upsilon(1S)\;\pi^+\pi^-, \Upsilon(2S)\;\pi^+\pi^-$, the rates 
are anomalously larger (by more than two orders of magnitude) than the expectations
from scaling the comparable $\Upsilon(4S)$ decays to the $\Upsilon(5S)$, and the shapes of the
distributions in the dipion invariant mass $m_{\pi \pi}$ and the cosine of the helicity angle,
$\cos \theta$, where $\theta$ is the angle between the $\pi^-$ and $\Upsilon(5S)$ in the dipion rest frame,
are not described by the models~\cite{Brown:1975dz}
based on the QCD multipole expansion~\cite{Gottfried:1977gp,Kuang:2006me} - 
a feature also at variance with similar dipion transitions between lower $\Upsilon$ resonances.
A critical observation towards understanding these features is that the final states in question
are produced not from the decays of $\Upsilon(5S)$, but from  the
process $e^+ e^- \to Y_b(10890) \to  \Upsilon(1S)\;\pi^+\pi^-, \Upsilon(2S)\;\pi^+\pi^-$,
with $Y_b$ a $1^{--}$ state, having a total decay width $\Gamma(Y_b)=55\pm 9$
 MeV~\cite{Olsen:2009gi}.
  In a closely related recent paper~\cite{Ali:2009pi}, we have analyzed the
BaBar data~\cite{:2008hx} obtained at the SLAC B factory during an energy scan of the
 $e^+e^- \to b \bar{b}$ 
cross section
 in the range of the center of mass energy $\sqrt{s}=10.54$ to 11.20 GeV, observing that the BaBar
 data on the $R_b$-scan are consistent with the presence of an additional $b\bar{b}$ state
 $Y_{[bq]}$ with a 
mass of 10.90 GeV and a width of about 30 MeV, apart from the $\Upsilon(5S)$ and $\Upsilon(6S)$
 resonances.
Identifying the $J^{PC}=1^{--}$  state $Y_{[bq]}(10900)$ seen in the energy scan of the
 $e^+ e^- \to b \bar{b}$
cross section by BaBar~\cite{:2008hx} with the state $Y_b(10890)$ seen by Belle~\cite{Abe:2007tk},
we present a dynamical model based on the tetraquark interpretation of $Y_b(10890)$
and show that it is in excellent agreement with the measured distributions in the
decays $Y_b \to  \Upsilon(1S)\;\pi^+\pi^-, \Upsilon(2S)\;\pi^+\pi^-$.

In the tetraquark interpretation,  $Y_{[bq]}$ is a $J^{PC}=1^{--}$ 
bound (diquark-antidiquark) state having the flavor content
 $Y_{[bq]}={\wQ}{\bar{\wQ}}=[bq][\bar{b}\bar{q}]$ (here $q=u$ or $q=d$, and $\wQ$ is a diquark) with the spin and orbital momentum
quantum numbers:
 $ S_{\wQ}=0,~S_{\bar{\wQ}}=0,~S_{\wQ\bar{\wQ}}=0,~L_{\wQ\bar{\wQ}}=1$~\cite{Drenska:2008gr}.
The first two quantum numbers are the
 diquark spin, antidiquark spin, respectively, and the last two denote the spin and
the orbital angular quantum numbers of the tetraquarks, with the total spin being
$J=S_{\wQ\bar{\wQ}} + ~L_{\wQ\bar{\wQ}}=1$. Such spin-0 diquarks are called ``good''
 diquarks~\cite{Jaffe:1976ih}
and an interpolating diquark operator can be written as
 ${\wQ}_{i\alpha}= \epsilon_{\alpha\beta\gamma}(\bar{b}_c^{\beta}\gamma_5 q_i^{\gamma}
-\bar{q}_{i_c}^\beta \gamma_5 b^\gamma)$ (with $q_i=u,d$ for
$i=1,2$ and $\bar{b}_c$ the charge conjugate $b$-quark field 
$\bar{b}_c=-ib^{\rm T} \sigma_2\gamma_5$). The ``good'' diquark ${\wQ}_{i\alpha}$ 
is in the attractive anti-triplet ($\bar{3}$) color channel (with the color quantum numbers denoted
by the Greek letters).  There are two such $J^{PC}=1^{--}$  states,
 $Y_{[bq]}=([bq]_{S=0} [\bar{b}\bar{q}]_{S=0})_{\rm P-wave}$,
 with the mass
 eigenstates, called $Y_{[b,l]}$ and $Y_{[b,h]}$ in
~\cite{Ali:2009pi},  being orthogonal combinations of $Y_{[bu]}$ and $Y_{[bd]}$. Their mass
 difference is induced by isospin splitting
 $m_d - m_u$ and a mixing angle and is estimated as $\Delta M(Y_b)=(5.6 \pm 2.8)$ MeV.
In the following, we will not distinguish between the lighter and the heavier of these states and denote
them by the common symbol $Y_b$.
The decays 
$
Y_b
\to \Upsilon(1S)\;\pi^+\pi^-,
\Upsilon(2S)\;\pi^+\pi^-$ are sub-dominant, but Zweig-allowed and involve essentially the quark rearrangements
 shown below.

With the $J^{PC}$ of the $Y_b$ and $\Upsilon(nS)$ both $1^{--}$, the $\pi^+\pi^-$ states
in the decays $Y_b \to \Upsilon(1S)\; \pi^+\pi^-, \Upsilon(2S)\; \pi^+\pi^-$ are allowed to have
the $0^{++}$ and $2^{++}$ quantum numbers. There are only three low-lying
 states in the Particle Data Group
 (PDG)~\cite{Amsler:2008zzb}  which
can contribute as intermediate states, namely the  two $0^{++}$ states, $f_0(600)$ and $f_0(980)$,
 which,
 following~\cite{Hooft:2008we,Maiani:2004uc},
we take as the lowest tetraquark states, and the $2^{++}$ $q\bar{q}$-meson state $f_2(1270)$, all of which decay dominantly
into $\pi \pi$. For the decay $Y_b \to \Upsilon(1S)\; \pi^+\pi^-$, all three states contribute. However,
kinematics allows only the $f_0(600)$ in the decay $Y_b \to \Upsilon(2S)\; \pi^+\pi^-$. In addition, 
a non-resonant contribution with a significant D-wave fraction is required by the data on
$Y_b \to \Upsilon(1S)\; \pi^+\pi^-, \Upsilon(2S)\; \pi^+\pi^-$ 
. The dynamical model described below encodes all these features.

 We start by showing the relevant diagrams for the decays
$Y_b(q) \to \Upsilon (p) + \pi^+(k_1) + \pi^-(k_2)$.

\begin{equation}  \label{verticesdef1}
\mathbf{a)}
\raisebox{-30pt}{\includegraphics[width=0.18%
\textwidth]{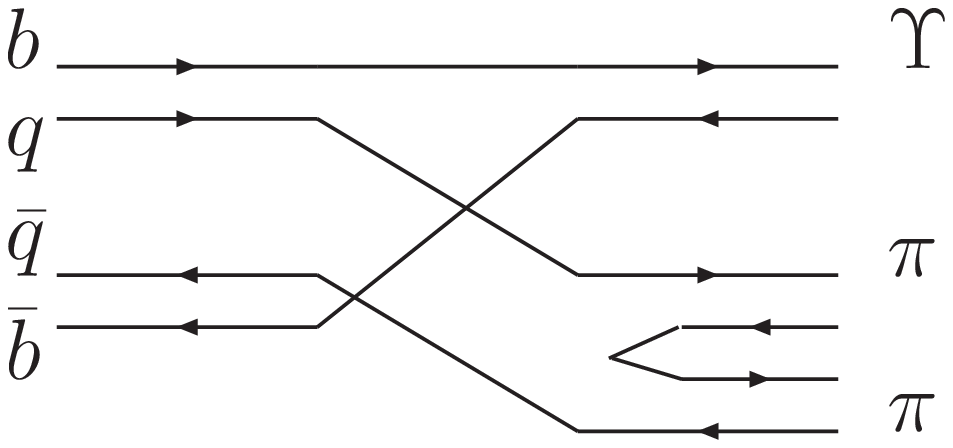}}
\en\en\en
\mathbf{b)}
\raisebox{-30pt}{\includegraphics[width=0.18%
\textwidth]{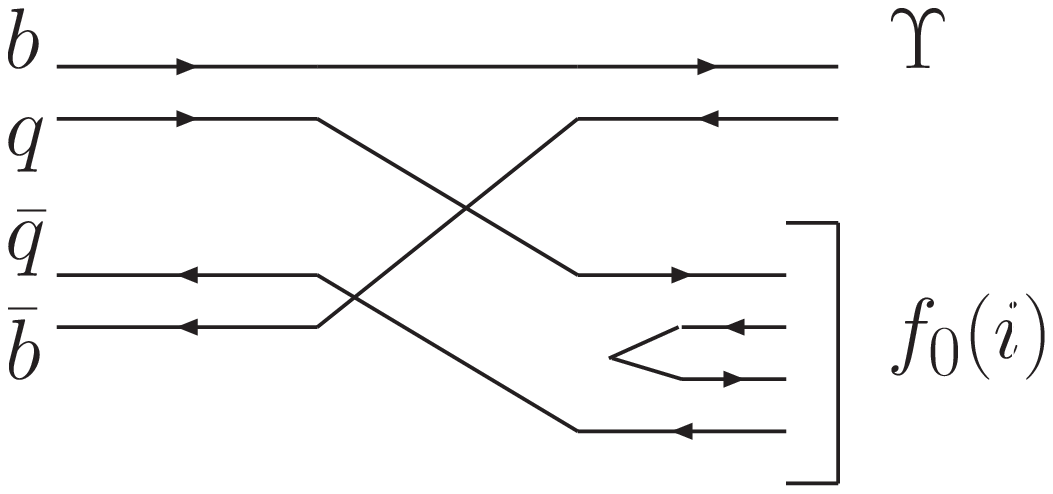}}
\end{equation}
The initial state represents the tetraquark states $Y_b=[bq][\bar{b}\bar{q}]$, and $\Upsilon$ stands
for $\Upsilon(1S)$ and $\Upsilon(2S)$. 
Both  diagrams involve the creation of a $q\bar{q}$ pair from the vacuum, with diagram $\mathbf{a}$
resulting into the (non-resonant) final states $\Upsilon(1S)\;\pi^+\pi^-$ and $\Upsilon(2S)\;\pi^+\pi^-$,
and diagram $\mathbf{b}$ leading to the final states 
$\Upsilon(1S)\; \left(f_0(600),f_0(980)\right)$ and
$\Upsilon(2S)\; f_0(600)$, with the implied subsequent decays $ \left(f_0(600),f_0(980)\right)\to \pi^+\pi^-$. The $2^{++}$
intermediate state $f_2(1270)$ contributing to the decay $Y_b \to \Upsilon(1S)\; \pi^+\pi^-$ is
depicted below. 
\begin{equation}  \label{verticesdef2}
\mathbf{c)}\raisebox{-30pt}{\includegraphics[width=0.26%
\textwidth]{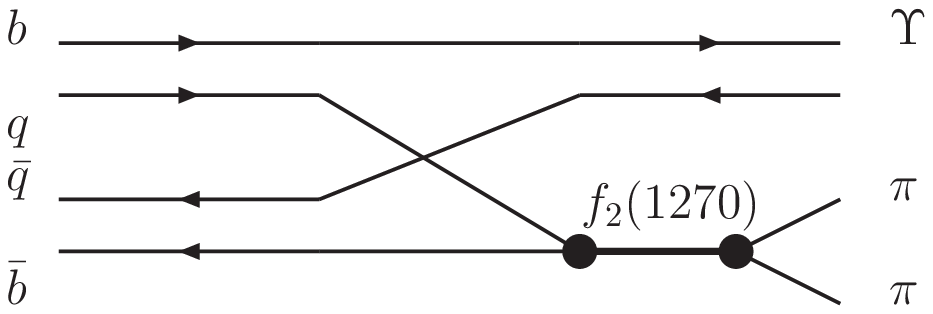}}
\end{equation}   

 Writing the Lorentz-invariant amplitudes as
\begin{equation}
 \mathcal{M} = \varepsilon^{Y}_\mu(q) \varepsilon^{\Upsilon}_\nu(p) \sum_{i=a,b,c}\mathcal{M}_{i}^{\mu\nu}(p,k_1,k_2)~, 
\end{equation}
where $\varepsilon^{Y}_\mu(q)$ and $\varepsilon^{\Upsilon}_\nu(p)$ are the polarization vectors of the
$Y_b$ and $\Upsilon(nS)$, respectively, we give below the explicit expressions
for $ \mathcal{M}_{i}^{\mu\nu}(p,k_1,k_2) $.

 The amplitude corresponding to  the non-resonant part $\mathbf{a})$
   is written, following Novikov and Shifman
 in~\cite{Brown:1975dz}, as
\begin{equation}  \label{verticesdef23}
\begin{array}{c}
\raisebox{-15pt}{\includegraphics[width=0.18%
\textwidth]{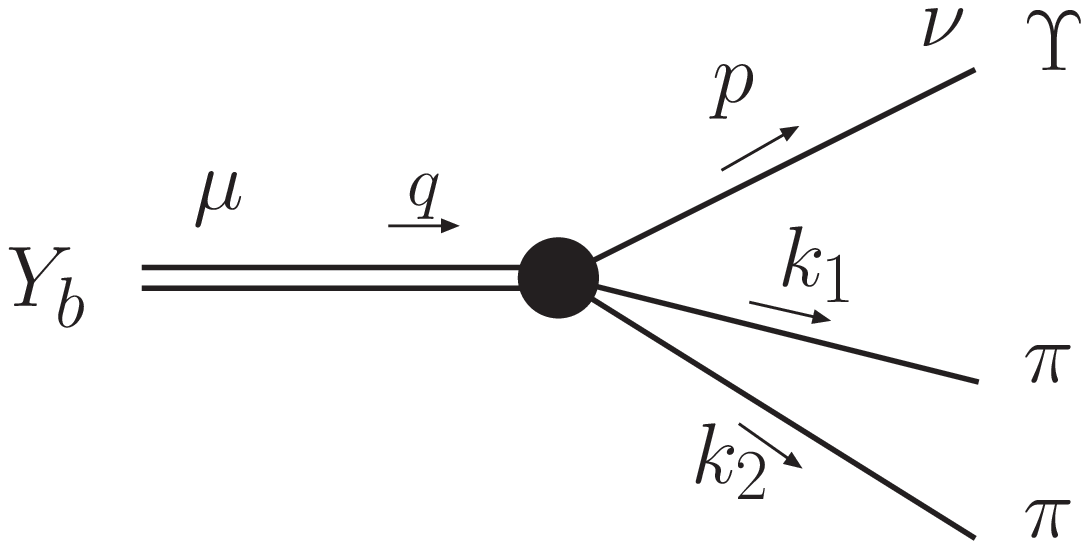}}  \widehat{=} 
\\ 
\mathcal{M}_{\mathbf{a}}^{\mu\nu}=g^{\mu\nu}\frac{F}{F_{\pi}^2} \textnormal{\LARGE{$[$}} 
m_{\pi\pi}^2-\beta (\Delta M)^2(1+\frac{2 m_{\pi}^2}{m_{\pi\pi}^2})+
\\
\frac{3}{2}\beta ((\Delta M)^2-m_{\pi\pi}^2)(1-\frac{4 m_{\pi}^2}{m_{\pi\pi}^2})(\cos^2 \theta -\frac{1}{3})\textnormal{\LARGE{$]$}} ,
\end{array}
\end{equation}    
Here $\Delta M = M_{Y_b}-M_{\Upsilon}$,  $F_{\pi}=130$ MeV is the pion decay constant, $m_{\pi \pi}=\sqrt{(k_1+k_2)^2}$ is the invariant mass of the two outgoing pions, 
 and $\theta$ is the angle between
the $\pi^-$ and $Y_b$ in the dipion rest frame. Eq.~(\ref{verticesdef23}) is a guess to model the $\pi\pi$
continuum, inspired by the  decay characteristics of the dipionic transitions involving Quarkonia
 states~\cite{Brown:1975dz},
such as  $\Upsilon(4S) \to \Upsilon(1S) \pi^+\pi^-$, in which the dipion mass spectra
do not show any resonant contributions. However, as we show here,
the dynamical quantities $F$
(a form factor) and $\beta$ (a measure of D-wave contribution) required to fit the data from 
the decays $Y_b \to \Upsilon(1S, 2S) \pi^+\pi^-$ are very different in magnitude from those 
required in the decay $\Upsilon(4S) \to \Upsilon(1S) \pi^+\pi^-$~\cite{:2009zy}.

 The amplitude $\mathcal{M}_{{\mathbf{b}}}^{\mu\nu}$ coming from  the diagram $\mathbf{b})$ is the  resonant part involving the
$0^{++}$ states $f_0(600)$ and $f_0(980)$, and the subsequent decays $f_0(600), f_0(980) \to \pi^+\pi^-$ :
\begin{equation}
\mathcal{M}_{\mathbf{b}}^{\mu\nu}= \frac{\Ff{i} F_{\pi} g^{\mu\nu} g_{\fz{i}} k_1. k_2 }
{k^2 -m_{\fz{i}}^2+ i m_{\fz{i}}  \Gamma_{\fz{i}}(m_{\pi \pi})}~,
\end{equation}
where $\fz{i}$ are the two $0^{++}$ resonances
and the various dynamical factors are defined below in terms of the relevant vertices and the propagator:
\begin{equation}  \label{verticesdef24}
\begin{array}{rcl}
\raisebox{-10pt}{\includegraphics[width=0.18%
\textwidth]{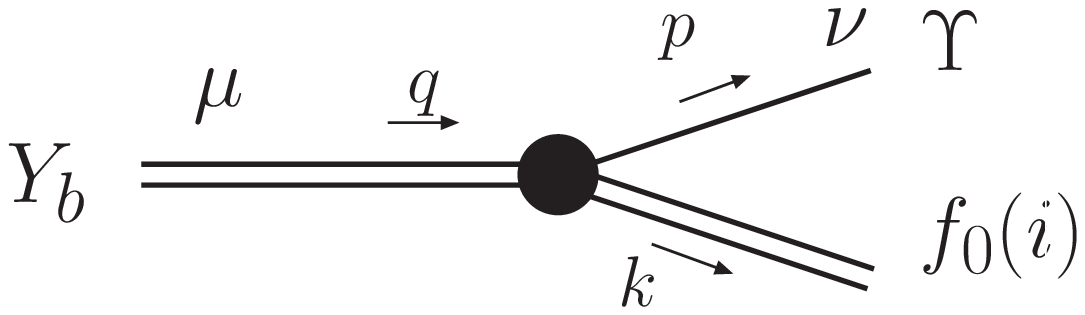}} & \widehat{=} & 
\Ff{i} F_{\pi} g^{\mu\nu}\;,  
\\
\raisebox{-10pt}{\includegraphics[width=0.14%
\textwidth]{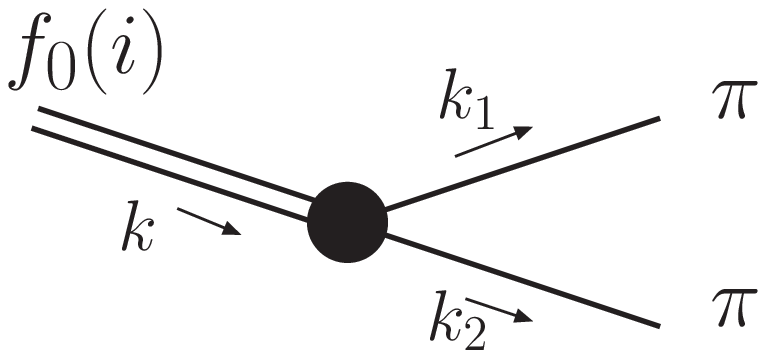}} & \widehat{=} &
g_{\fz{i}} k_1. k_2\;,
\\
\raisebox{-0pt}{\includegraphics[width=0.10%
\textwidth]{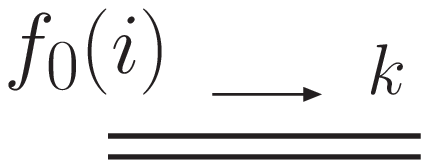}} & \widehat{=} &
\frac{1}{k^2 -m_{\fz{i}}^2+ i m_{\fz{i}}  \Gamma_{\fz{i}}(m_{\pi \pi})}\;,\en\en
\end{array}
\end{equation}    
and $\fz{i}=\fz{600}$ or $\fz{980}$. The couplings $g_{f_0(600)}=-c_f$ and $g_{f_0(980)} =\sqrt{2} c_I$ are taken
 from \cite{Hooft:2008we}, where $c_f=0.02\pm 0.002~{\mbox MeV}^{-1}$ and
 $c_I=-0.0025\pm0.0012~{\mbox MeV}^{-1}$. We use the central values for the couplings.
The propagator of $\fz{600}$ should not be taken in the minimal width approximation, since the total
 decay width and
 the mass are of the same order~\cite{Amsler:2008zzb,Caprini:2005zr}. Following \cite{Aitala:2000xu},
 the width is multiplied by a momentum-dependent factor:
\begin{equation}
\Gamma(m_{\pi \pi})=\Gamma_{\fz{600}} \frac{m_{\fz{600}}}{m_{\pi \pi}} \frac{p^*}{p_0^*},
\end{equation}
where $p_0^*=p^*(m_{\fz{600}})$ and $p^*=p^*(m_{\pi \pi})$ are the decay momenta  in the
 resonance rest frame. 
The other scalar ($f_0(980)$), having $\Gamma_{f_0(980)}/m_{f_0(980)} \ll 1$, 
is taken in the minimal width approximation, i.e. $\Gamma(m_{\pi \pi})=\Gamma_{\fz{980}}$. 

The amplitude $\mathcal{M}_{\mathbf{c}}^{\mu\nu}$ coming from  diagram $\mathbf{c})$ is 
\begin{eqnarray}
\mathcal{M}_{\mathbf{c}}^{\mu\nu}&=&g^{\mu\nu}A_{f_2(1270)}(m_{\pi\pi})=
g^{\mu\nu}\frac{\sqrt{8\pi (2J +1)}}{\sqrt{ m_{\pi\pi}}} Y_2^2 
\nonumber \\&&
\times \frac{a_{f_2(1270)}  \sqrt{ m_{f_2(1270)}}}
{m_{f_2(1270)}^2-m_{\pi\pi}^2- i m_{f_2(1270)}\Gamma_{f_2(1270)}}.
\end{eqnarray}
For $f_2(1270)$, $J=2$, and  we have kept only the helicity-2 component of the D wave with $Y_2^2$ the
corresponding spherical harmonics, $|Y_2^2|=\sqrt{\frac{15}{32\pi}}  \sin^2 \theta$. In principle, there
is also a helicity-0 component of the D wave $Y_2^0$ present in the amplitude, but 
following the high statistics experimental measurement of the process
 $\gamma \gamma \to f_2(1270) \to \pi^+\pi^-$ by Belle~\cite{Uehara:2008pf}, this contribution is small,
characterized by the value of $r_{02}$, the helicity 0-to helicity 2 ratio in $f_2(1270) \to \pi \pi$,
$r_{02}=(3.7 \pm 0.3 ^{+15.9}_{-2.9})\%$.  This can be included as more precise measurements become
 available.   

The described diagrams yield a coherent amplitude, and the various
 contributions  interfere with each other
 having non-trivial strong (interaction) phases, which are
{\it a priori} unknown. We treat them as free parameters to be determined by the fits to
 the Belle data. Combining all three amplitudes, the complete decay amplitudes for
 $Y_b \to \Upsilon(1S)\; \pi^+\pi^-, \Upsilon(2S)\; \pi^+\pi^-$ are:
\begin{eqnarray}
\label{model}
\mathcal{M}&=&
\varepsilon^{Y} . \varepsilon^{\Upsilon}
\textnormal{\LARGE{$[$}}
\sum\limits_{res}
\raisebox{-15pt}{\includegraphics[width=0.1%
\textwidth]{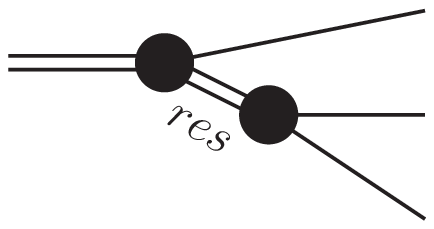}}+\raisebox{-15pt}{\includegraphics[width=0.1%
\textwidth]{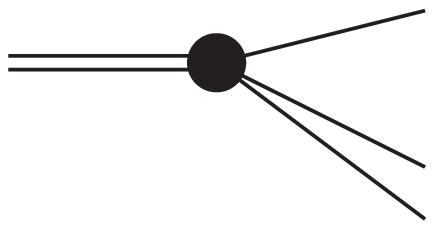}} 
\textnormal{\LARGE{$]$}}
\nonumber\\
&=&\varepsilon^{Y} . \varepsilon^{\Upsilon}
\textnormal{\Huge{$[$}}
 \frac{F}{F_{\pi}^2} 
 \textnormal{\LARGE{$[$}}
 m_{\pi\pi}^2-\beta (\Delta M)^2(1+\frac{2 m_{\pi}^2}{m_{\pi\pi}^2})+
\nonumber\\&&
\frac{3}{2}\beta((\Delta M)^2-m_{\pi\pi}^2)(1-\frac{4 m_{\pi}^2}{m_{\pi\pi}^2})(\cos^2 \theta -\frac{1}{3})\textnormal{\LARGE{$]$}}
\nonumber\\&&
+\sum\limits_{i}
\frac{a_{\fz{i}} e^{i\varphi_{\fz{i}}}  (m_{\pi\pi}^2-2 m_\pi^2)/2}{m_{\pi\pi}^2 -m_{\fz{i}}^2+ i m_{\fz{i}}  \Gamma_{\fz{i}}(m_{\pi\pi})}
\nonumber\\&&+
a_{f_2(1270)} e^{i\varphi_{f_2(1270)}}  A_{f_2(1270)}(m_{\pi\pi})
\textnormal{\Huge{$]$}},
\end{eqnarray}
where $a_{\fz{i}}=g_{\fz{i}}\Ff{i} F_{\pi}$. 
The sum over $i$ runs over all $0^{++}$ resonances contributing in the given energy range.

The differential decay width (averaged over the polarizations of the initial $Y_b$-hadron
 and
 summed over polarizations of the final $\Upsilon$-meson) is given by
\begin{equation}
d\Gamma=\frac{1}{(2\pi)^3}\frac{1}{32 M_{Y_b}^3}\overline{|\mathcal{M}|^2} dm_{\Upsilon \pi}^2 dm_{\pi \pi}^2,
\end{equation}
where $m_{\Upsilon \pi}^2=(p+k_1)^2$ (the amplitude is symmetric under the interchange of the two pions).
The $\cos \theta$ dependence is given implicitly by $m_{\Upsilon \pi}$.
By integrating over the phase space, we derive the two distributions in $m_{\pi \pi}$ and 
 $\cos \theta $. 

\begin{figure}[H]
\centering
\includegraphics[width=0.48\textwidth]{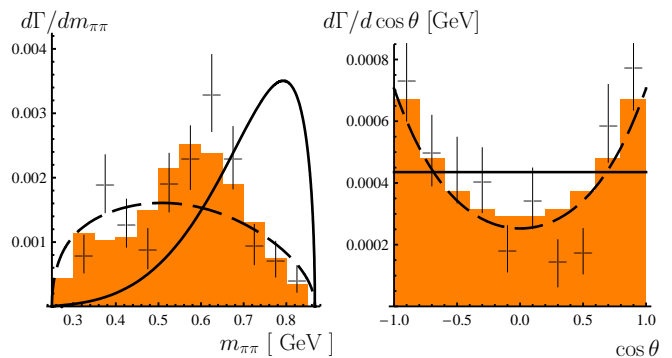} 
\vspace{-18pt}
\caption{
Dipion invariant mass ($m_{\pi\pi}$) distribution (left frame) and the $\cos \theta$
distribution (right frame) measured by  Belle~\cite{Abe:2007tk} for the final state
$\Upsilon(2S) \pi^+\pi^-$ (crosses), and the theoretical distributions based on this work
(histograms). 
The solid and dashed lines show purely continuum contributions for different $\beta$.
 \label{fitvalues1res2s}
}
\end{figure}

\begin{figure}[H]
\centering
\includegraphics[width=0.48\textwidth]{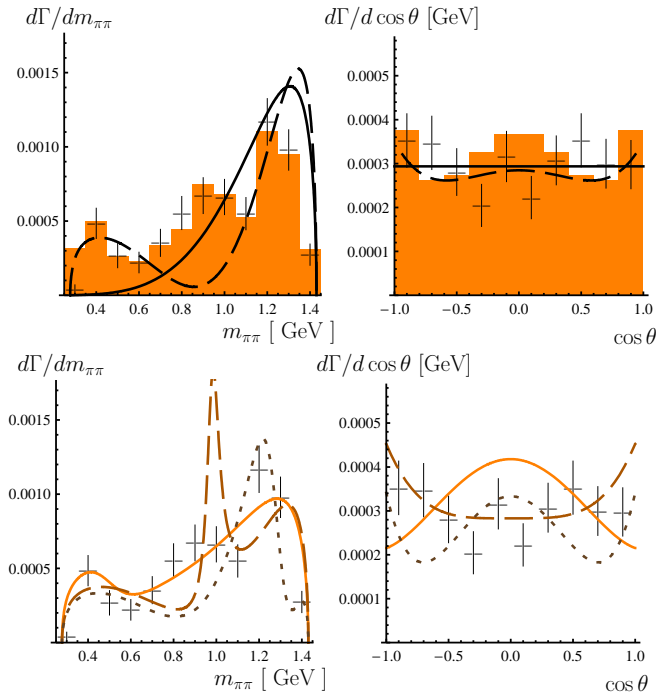} 
\vspace{-18pt}
\caption{ 
Upper Frames: The distributions measured by Belle~\cite{Abe:2007tk} for the final state
$\Upsilon(1S) \pi^+\pi^-$ (crosses), and the theoretical distributions based on this work (histograms).
The solid and dashed lines show purely continuum contributions for different $\beta$.
Lower Frames: Contributions with continuum plus a single resonance (solid curves: $f_0(600)$; dashed curves:
 $f_0(980)$; dotted curves: $f_2(1270)$).
}
\label{fitvalues1res1s}
\end{figure}

We have undertaken  fits of the Belle data \cite{Abe:2007tk} with our model
(\ref{model}), normalizing the distributions for the
 $\Upsilon(1S)\pi^+\pi^-$ and $\Upsilon(2S)\pi^+\pi^-$ channels to yield the measured partial decay widths
 $\Gamma_{\Upsilon(1S)+2\pi}=0.59\pm 0.04\pm 0.09~{\mbox MeV}$ and
 $\Gamma_{\Upsilon(2S)+2\pi}=0.85\pm 0.07 \pm 0.16~{\mbox MeV}$. The input parameters given in
 Table~\ref{tt} are taken from the PDG~\cite{Amsler:2008zzb}, except for the $\fz{600}$, for which we have
 taken the values from E791~\cite{Aitala:2000xu} .
\begin{table}[H]
\caption{Input masses and decay widths (in GeV) of the resonances $\fz{600}$, $\fz{980}$ and
$f_2(1270)$.}
\label{tt}
\begin{center}
\vspace{-7pt}
\begin{tabular}{|l|l||l|l||l|l|}
\hline
$ M_{Y_b} $ & $ 10.890  $ &$m_{f_0(600)} $ & \mfzsixh  & $\Gamma_{f_0(600)}$ & \Gfzsixh  
\\ \hline
$ M_{\Upsilon(1S)} $ & $ 9.460 $ &$m_{f_0(980)}$ & 0.980 & $\Gamma_{f_0(980)}$ & 0.07  
\\ \hline
$ M_{\Upsilon(2S)} $ & $ 10.023 $ &$m_{f_2(1270)}$ & 1.270 & $\Gamma_{f_2(1270)}$ & 0.185 
\\ \hline\hline
\end{tabular}%
\end{center}
\end{table}
\vspace{-14pt}
The dipion invariant mass distribution $d \Gamma/dm_{\pi \pi}$ and the angular 
distribution $d \Gamma/d \cos \theta$ [GeV]
 measured by Belle~\cite{Abe:2007tk} for the final
 state $\Upsilon(2S) \pi^+\pi^-$
are shown in Fig. \ref{fitvalues1res2s}. 
The shaded histograms are the
corresponding theoretical  distributions from our model having a
 $\chi^2/{\rm d.o.f.} \approx 9/8$ (obtained for the $m_{\pi\pi}$ spectrum),
 with the fit parameters given in
 Table~\ref{fitvaluest1res1s}, yielding an
 integrated decay width of
$\Gamma(Y_b \to \Upsilon(2S) \pi^+\pi^-)=0.85$ MeV. 
The solid curves are the distributions for
 $\beta=0$  from the non-resonant
part \eqref{verticesdef23}  alone, which are the anticipated distributions
from the decays $\Upsilon(5S) \to \Upsilon(2S) \pi^+\pi^-$ ~\cite{Brown:1975dz,:2009zy}. The dashed curves
correspond to the best-fit solution without the $f_0(600)$ contribution, yielding $\beta\approx 0.4$ with $\chi^2/d.o.f \approx 23/10$ (obtained for the $m_{\pi\pi}$ spectrum). 
The difference between the histograms (our fits) and the curves is that the latter do not have the
$f_0(600)$ contribution. Both the solid and dashed curves fail to describe the Belle data.
\vspace*{-8mm}
\begin{table}[H]
\caption{Fit values, yielding $F=0.86\pm 0.34$, $\beta=0.7\pm 0.3$
for the non-resonant contribution, and for the parameters entering in the resonant
 amplitude from $\fz{600}$ for the decay $Y_b \to \Upsilon(2S) \pi^+\pi^-$. } \
\label{fitvaluest1res1s}
\begin{center}
\vspace{-14pt}
\begin{tabular}{|l|l|l|l|l|}
\hline
 & $a_{f_0(i)}$   &  $F_{f_0(i)}$  & $F_{f_0(i)}/F$ & \
$\varphi_{f_0(i)}$ (rad.) \\ \hline
$f_0(600)$& $10.89\pm 2.4$ & $4.19\pm 0.92$ & $4.86\pm 2.18$ & \
$2.76\pm 0.22$ \\ \hline\hline
\end{tabular}%
\end{center}
\end{table}
\vspace{-14pt}
The measured spectra (in $m_{\pi\pi}$ and $\cos \theta$) for the final state  $\Upsilon(1S)\pi^+\pi^-$
from Belle~\cite{Abe:2007tk} are shown in 
Fig.~\ref{fitvalues1res1s} together with our theoretical distributions (histograms) obtained for the model in \eqref{model}
having a $\chi^2/{\rm d.o.f.} \approx 5/5$ (obtained for the $m_{\pi\pi}$ spectrum in the upper left frame), with the fit parameters given in Table~\ref{fitvaluest3res1s} yielding an integrated decay width of
$\Gamma(Y_b \to \Upsilon(2S) \pi^+\pi^-)=0.66$ MeV.
The two curves in the upper frames show the shape of the
 continuum contribution based on \eqref{verticesdef23}, with the solid curves obtained for
 $\beta=0$ (as would be expected for
 the transition $\Upsilon(5S) \to \Upsilon(1S) \pi^+\pi^-$),
 and the dashed curves
corresponding to the best-fit solution without the resonant contributions yielding $\beta\approx 0.3$ with $\chi^2/d.o.f \approx 65/11$ (obtained for the $m_{\pi\pi}$ spectrum). Both of them fail to describe the Belle data.
In addition we show the contributions from the continuum plus a single resonance in the lower frames (solid curves: $f_0(600)$ with $\chi^2/d.o.f \approx 16/9$; dashed curves:
 $f_0(980)$ with $\chi^2/d.o.f \approx 30/9$; dotted curves: $f_2(1270)$ with $\chi^2/d.o.f \approx 33/9$).
 They also fail to describe the Belle data.
\vspace*{-8mm}
\begin{table}[H]
\caption{Fit values, yielding $F=0.19\pm 0.03$,   $\beta=0.54\pm 0.12$
for the non-resonant contribution, $a_{f_2(1270)}=0.5\pm 0.16$, $\varphi_{f_2(1270)}=3.33\pm 0.06$ for $f_2(1270)$, and for the parameters entering in the resonant amplitude from
$\fz{600}$ and $\fz{980}$ for the decay $Y_b \to \Upsilon(1S) \pi^+\pi^-$.}
 \label{fitvaluest3res1s}
\begin{center}
\vspace{-7pt}
\begin{tabular}{|l|l|l|l|l|}
\hline
 & $a_{f_0(i)}$   &  $F_{f_0(i)}$ &  $F_{f_0(i)}/F$ &  $\varphi_{f_0(i)}$ (rad.) \\ \hline
$f_0(600)$&$3.6\pm 0.7$ & $1.38\pm 0.27$ & $7.34\pm 1.94$ & $1.14\pm  0.14$  \\ \hline  
$f_0(980)$&$0.47\pm 0.02$ & $1.02\pm 0.04$ & $5.42\pm 1.0$ & $4.12\pm  0.3$ \\ \hline  
\hline
\end{tabular}%
\end{center}
\end{table}
\vspace{-14pt}

We also remark that using the fits of the data for the decay $Y_b \to \Upsilon(1S) \pi^+\pi^-$
presented here, we are able to explain the decay width for the decay $Y_b \to \Upsilon(1S) K^+ K^-$,
measured by Belle~\cite{Abe:2007tk}. The decay is anticipated to
be strongly dominated by the $0^{++}$ tetraquark state $f_0(980)$. Details will be published elsewhere.

 Summarizing, we have argued here that the decays $Y_b \to \Upsilon(1S,2S) \pi^+\pi^-$
are radically different than the similar dipion transitions measured in the $\Upsilon(4S)$ and
lower mass Quarkonia. 
The dynamical model presented by us will be tested in great detail with improved data, which we
expect in the near future from Belle.

We thankfully acknowledge helpful communications with the members of the Belle and BaBar
 collaborations and thank Alexander Parkhomenko for his comments.
M.J.A. would like to thank DESY for the hospitality and
ENSF, Trieste, for financial support.

\end{document}